\documentclass[aps,prd,twocolumn,superscriptaddress,nofootinbib]{revtex4}

\usepackage{epsfig}
\usepackage{amssymb}
\usepackage{amsmath}
\usepackage{amsfonts}
\usepackage{hyperref}

\newcommand{\be}{\begin{equation}}
\newcommand{\ee}{\end{equation}}
\newcommand{\bea}{\begin{eqnarray}}
\newcommand{\eea}{\end{eqnarray}}
\newcommand{\bi}{\begin{itemize}}
\newcommand{\ei}{\end{itemize}}
\newcommand{\ben}{\begin{enumerate}}
\newcommand{\een}{\end{enumerate}}

\newcommand{\ep}{\epsilon)}

\def\gsim{\mathrel{\rlap{\lower4pt\hbox{\hskip1pt$\sim$}}
    \raise1pt\hbox{$>$}}}         
\def\lsim{\mathrel{\rlap{\lower4pt\hbox{\hskip1pt$\sim$}}
    \raise1pt\hbox{$<$}}}         

\def \ep {\epsilon}

\begin{document}

\title{On the NNLO QCD corrections to single-top production at the LHC}

\author{Mathias Brucherseifer}
\email{mathias.brucherseifer@kit.edu}
\affiliation{Institute for Theoretical Particle Physics,
Karlsruhe Institute of Technology, Karlsruhe, Germany}
\author{Fabrizio Caola}
\email{caola@pha.jhu.edu}
\affiliation{Department of Physics and Astronomy, Johns Hopkins 
University, Baltimore, USA}
\author{Kirill Melnikov}
\email{melnikov@pha.jhu.edu}
\affiliation{Department of Physics and Astronomy, Johns Hopkins 
University, Baltimore, USA}

\begin{abstract}
 We present a fully-differential calculation of the NNLO QCD corrections to the  
$t$-channel mechanism for producing  single top quarks at the LHC. 
We work in the structure 
function approximation, computing  QCD corrections to the
light- and heavy-quark lines separately and neglecting  the 
dynamical cross-talk between the two. 
The neglected contribution, which appears at NNLO for the first time, 
is color-suppressed and is expected to be sub-dominant. 
Within this approximation, we find that, for the total cross section, 
NNLO QCD corrections  are in the few percent range and, therefore, 
are comparable to NLO QCD corrections. We also find that the scale 
independence of the theoretical prediction for  single-top production 
improves significantly once  NNLO QCD corrections are included.  Furthermore, we show how these results 
change if a cut on the transverse momentum of the top quark is  applied and derive the NNLO QCD prediction 
for the ratio of single top and single anti-top production cross sections at the 8 TeV LHC. 
\end{abstract}

\maketitle

\section{Introduction}

Studies of top quarks produced in hadron collisions  are important for understanding many properties 
of these heavy particles, including their masses,  their  couplings to electroweak 
gauge bosons, their Cabbibo-Kobayashi-Maskawa matrix element $V_{tb}$ etc. 
In many cases, the precision reached in measuring these quantities is already close to a few a percent, thanks 
to the successful top quark physics programs at the Tevatron and the LHC. 
Further high-statistics data samples, that will become available during a forthcoming 
$13~{\rm TeV}$ run of the LHC, will  remove statistical uncertainties as a limiting factor for these 
measurements (see e.g.~\cite{Agashe:2013hma}).
As the result, theoretical uncertainties related to imprecise knowledge 
of  production cross sections and kinematic distributions will become 
an important limiting factor in pushing precision measurements forward.  

There are two main mechanisms for producing top quarks in  hadron collisions. 
Both at the Tevatron and the LHC, the dominant one occurs due to 
strong interactions and, through such processes as $q \bar q \to t  \bar t$ or $gg \to t \bar t$,  leads 
to the production of $t \bar t$ pairs. The theoretical description of 
this production mechanism is very advanced; it includes NLO QCD and electroweak corrections, 
soft gluon resummations and, since recently, complete NNLO QCD 
corrections~\cite{Cacciari:2011hy,Beneke:2009rj,Czakon:2009zw,Bonciani:1998vc,Kuhn:2006vh,Czakon:2013goa}.
 The second mechanism is governed  by weak interactions and relies on 
the flavor changing transitions  $W^* \to tb$, $b \to t W $ or $W^* b \to t$  to produce single top (or anti-top) quarks. 
Although sub-dominant relative to  $t \bar t$ pair production, this mechanism yields a 
sizable fraction of top quark events both at the Tevatron and the LHC. 
Experimental  conditions for studying single-top production at the two colliders 
 are however,  very different. 
Indeed, when top quarks decay, they produce leptons, missing energy and $b$-jets. Correspondingly, the
main  background  for observing single-top production at a hadron collider  is the direct production 
of  $W$ bosons in association with jets in general and with $b$-jets in particular.   
The severity of this background 
and the relative smallness ${\cal O}(1~{\rm pb})$ 
of the  single-top production cross section  made detailed studies 
of this process at the Tevatron very difficult.  Nevertheless, 
the CDF and D0 collaborations confirmed the existence of 
the electroweak production mechanism  for top quarks 
 and measured the cross section for this process with, approximately, twenty percent 
precision~\cite{stcdf,std0,Abazov:2011pt,cdfconf}.   Since the single-top production 
cross section is proportional  to  the electroweak coupling of a top quark  
to a $W$-boson, a 
${\cal O}(20\%)$ measurement of the production cross section can be interpreted as a  ${\cal O}(10\%)$ 
measurement of the CKM matrix element $|V_{ tb}|$ or a ${\cal O}(20\%)$ measurement of the top quark width. 

Experimental conditions improve dramatically  at the LHC where the single top quark production cross section 
is significantly higher, approximately $60~{\rm pb}$ at the $8~{\rm TeV}$ LHC and $160~{\rm pb}$ at the $14~{\rm TeV}$ 
LHC.   Given that expected integrated LHC luminosities are in the range of a few hundreds inverse femtobarns, 
millions of top quarks will be produced at the LHC 
by virtue of electroweak interactions alone, making  high-precision studies of this production mechanism 
 an important part of the experimental program.  Indeed, already in the first run of the LHC, ATLAS and CMS collaborations 
improved significantly on the CDF and D0 results,  by measuring the single-top -production cross sections with a ten 
percent accuracy~\cite{statlas,stcms,Aad:2012xca,Chatrchyan:2014tua,ATLASandCMSCollaborations:2013ofa}.
Similar to what we discussed in the context of the Tevatron, such a measurement 
can be interpreted as
a ${\cal O}(5\%)$ measurement of $|V_{tb}|$ 
and a ${\cal O}(10\%)$ measurement of the top quark width. 
This is the highest experimental precision available for these 
quantities currently. 

It is important to emphasize that  there  are several experimentally distinguishable  ways 
to produce single  top quarks through electroweak interactions. 
 Indeed, writing  the  primary electroweak 
$tbW$-vertex  in three different  ways, $W^* b \to t$ , $W^* \to tb$, $b \to t W $,  we obtain distinct mechanisms 
for single top quark production that are usually referred to as the 
$t$-channel ($W^* b \to t$) process, the  $s$-channel process ($ W^* \to tb $) and the 
$tW$ production ($ b \to Wt $). Among these three mechanisms, the $t$-channel process has the largest 
cross section both at the $8$ TeV LHC and at the Tevatron contributing, respectively, $82\%$ and $65\%$ to  the total  
cross section $\sigma_t$.   The $s$-channel 
process is $33\%$ of $\sigma_t$ at the Tevatron and is about $5\%$ at the $8$ TeV  LHC. The $tW$ production 
is negligible at the Tevatron and contributes ${\cal O}(15\%)$ 
to $\sigma_t$ at the $8$ TeV LHC. 
However, since $tW$ production  can be distinguished 
from the other two mechanisms, it is usually treated separately in experimental analyses. 
Also, we note that  for the higher-energy LHC, the $t$-, $s$- and $tW$ production channels contribute in 
similar proportions as for the $8$ TeV LHC.

\begin{figure*}[th]
\centering
\includegraphics[scale=0.4]{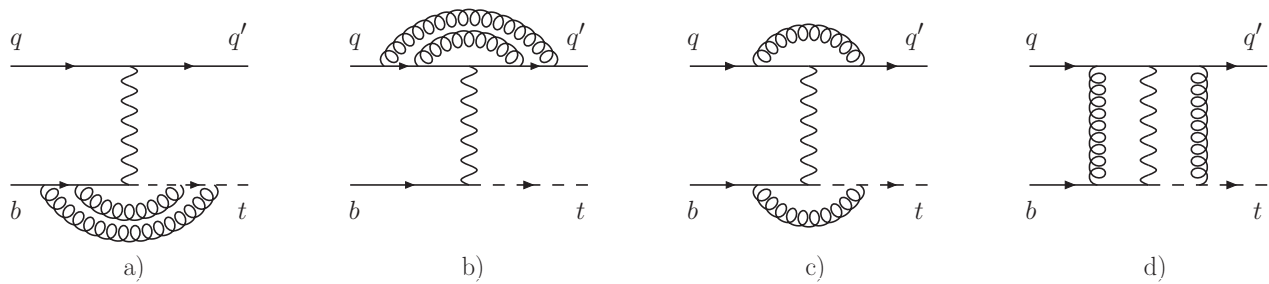}
\caption{Schematic representation of  different contributions to the
NNLO QCD corrections for the $t$-channel single-top production.
From left to right: 
a) NNLO corrections the the heavy quark line, b) NNLO
corrections to the light quark line, c) product of NLO corrections
to the  heavy and the light quark lines, and d) 
non-factorizable contributions that are neglected in 
this paper. Corresponding real emission diagrams, as well
as crossed channels, are not shown.
}\label{feynman_blocks}
\end{figure*}

Theoretical results for single top quark production are available at an ever increasing level of sophistication.
These include NLO QCD and electroweak predictions in four- or five-flavor scheme for both stable~\cite{nlo1,nlo2,nlo3,
Beccaria:2008av,Heim:2009ku}
and decaying~\cite{nlo4,nlo5,nlo6,nlo7,nlo8,Schwienhorst:2010je,
Pittau:1996rp,Falgari:2010sf,Falgari:2011qa}
  top quarks,   
resummations~\cite{Kidonakis:2011wy,Kidonakis:2010tc,Zhu:2010mr,Wang:2010ue}
 and  fixed order computations matched to parton showers~ \cite{mc1,mc2,
Frederix:2012dh,Alioli:2009je}.
Focusing on NLO QCD corrections, we note that  they are small, of the order 
of a few percent, for the $t$-channel single-top production. On the other hand, 
NLO QCD corrections for the  $s$-channel single-top production are large, 
of the order of fifty percent,   both at the Tevatron and the LHC. 
Corrections to the associated $tW$ production are known to be moderate at both colliders~\cite{nlo5}. 

We note that the   smallness of the NLO QCD corrections to the $t$-channel single-top production cross section 
is the result of strong cancellations  between different sources of such  corrections, e.g. different partonic 
channels.
It is unclear if these cancellations are accidental or generic and if the smallness of NLO QCD corrections 
implies that NNLO QCD corrections are, in fact,  even smaller as should be the case for convergent perturbative series.
The most obvious reflection 
of this fact is the  strong sensitivity of the NLO QCD prediction for $t$-channel single-top production 
to choices of factorization and renormalization 
scales. In fact, the sensitivity of the NLO QCD corrections to these scales is comparable to the size of 
corrections themselves.   Therefore, if the scale variation is an indication of the size of missing higher-order QCD
corrections,  it is tempting  to conclude that 
NLO QCD corrections to $t$-channel single-top production cross section  are {\it accidentally} 
small and that the natural size of NNLO QCD corrections to this process is at a few percent level.
Therefore,  we expect that NLO and NNLO QCD corrections to the single-top production cross section are of  a similar size. 
This implies that  NNLO QCD corrections to   
$t$-channel single-top production cross section  must be computed to enable studies of  electroweak 
production of top quarks at the   LHC  with a percent accuracy.

The goal of this paper is to make the first step towards a high-precision prediction for the single-top production 
at the LHC, by providing  fully-differential NNLO QCD corrections to  the $t$-channel single top quark production 
in the approximation where corrections to light quark $q \to q' W^*$ and heavy  quark $ W^* b \to t$  
weak
transitions are treated (almost) independently from each other.
More precisely,
we neglect all dynamical cross-talk between corrections to the light and heavy quark lines, which
then depend on each other only through kinematic phase-space constraints.
At NLO, this approximation is 
exact due to color conservation. At NNLO however, the exchange of 
two (real or virtual) gluons in a color-singlet
state between light and heavy quark lines, shown 
in Fig.~\ref{feynman_blocks}d, leads to a non-vanishing contribution to 
the cross section.   
We expect this contribution to be small since it is suppressed by at least 
two powers of the number of colors $N_c=3$ relative to 
the ``factorizable'' contributions shown in 
Fig.~\ref{feynman_blocks}a-c. Therefore, 
we neglect the non-factorizable contributions in the rest of the paper.

The paper is organized as follows. 
In Section~\ref{setup} we briefly discuss the technical details of the 
calculation.  In Section~\ref{results} we show some results for NNLO QCD corrections to single-top and single anti-top 
production at the $8$ TeV LHC.  We conclude in Section~\ref{concl}. 

\section{Technical details of the calculation}
\label{setup}

Our goal is to compute NNLO QCD corrections to $t$-channel  single  top quark production.
The top quarks are considered stable. In the approximation where only factorizable corrections 
are retained, 
the calculation can be divided into three different
parts. We need to compute {\it i})  NNLO QCD corrections to the weak transition on a heavy 
quark line  $W^* b \to t$, c.f. Fig.\ref{feynman_blocks}a;
{\it ii}) NNLO QCD corrections to the weak transition on a light quark line  $u \to W^* d$, c.f. 
Fig.\ref{feynman_blocks}b; and 
{\it iii}) a product of NLO QCD corrections to weak transitions on both heavy and light quark lines,
c.f. Fig.\ref{feynman_blocks}c.
These three contributions are individually
infra-red and collinear finite, gauge invariant, and can be considered separately. 
We now briefly illustrate some of the technical details of our computation, starting from corrections
to the heavy quark line. 

As a preliminary remark,  we note that the computation of NNLO QCD corrections to  a process $X$ 
requires  three  ingredients: 1) two-loop  QCD corrections 
to $X$; 2) one-loop QCD corrections to $X+$jet and 3) tree-level matrix element for the $X+2~$jet process.
All of the required ingredients to compute 
NNLO QCD corrections to the $W^*b \to t $ transition, for arbitrary invariant mass of the $W$-boson, 
 can be obtained by crossing the two-loop, 
one-loop and tree amplitudes used by 
us recently in the computation of NNLO QCD corrections to top quark 
decay $t \to W^* b$~\cite{Brucherseifer:2013iv}.
This crossing is straightforward for one- and two-loop virtual amplitudes to the $0 \to tW \bar b$ vertex~\cite{nnlov1,nnlov2,nnlov3,nnlov4,nnlov5} 
(since they depend on a very 
small number of kinematic invariants) 
and for tree-level amplitudes  $t \to b W^* gg$ and 
$t \to W^* b q\bar q$.  The crossing is potentially more 
challenging for one-loop corrections to $t \to b g W^*$~\cite{nlo5}, that we borrow from 
the MCFM program~\cite{mcfm} since, in this case, the number of kinematic invariants is larger. 
To ensure that this analytic continuation is correct,  we computed 
$0 \to t \bar b g W^*$ amplitudes in physical kinematics, where an off-shell $W^*$ boson 
and a  $b$-quark  collide to produce a gluon and a top quark, 
using our own implementation of the Passarino-Veltman 
reduction procedure for one-loop tensor integrals~\cite{Passarino:1978jh},
and found complete agreement with the result obtained 
by crossing one-loop MCFM amplitudes. 

In general, the computation of NNLO QCD corrections to {\it any} process is made complicated by the fact that all three 
ingredients for NNLO computations that we listed above are separately infra-red and collinear divergent. 
We regularize and extract these  divergences by constructing 
subtraction terms, following the  approach described in 
Refs.~\cite{Czakon:2010td,Czakon:2011ve,Boughezal:2011jf}\footnote{For other approaches to NNLO computations, see~\cite{nnlo_tech}. 
Some phenomenological applications can be found in~\cite{Czakon:2013goa,nnlo_pheno,Boughezal2013,Brucherseifer:2013iv}.}.
 The subtraction terms 
involve products of lower-multiplicity matrix elements  with appropriate splitting functions 
or eikonal currents that need to be computed for both initial and final  state partons.  The initial state 
singularities were absent in the computation of NNLO QCD corrections to top decay~\cite{Brucherseifer:2013iv}
but they are an important  part in the computation of single-top production. 
The corresponding splitting functions, both tree and one-loop, are well-known~\cite{lcat1,lb,lmc3,lcat2,lzb1,lcat3,lcat4,Kosower:1999rx}
and the one-loop 
eikonal current  for the $bW^* \to t$ transition~\cite{Bierenbaum:2011gg}  is  again obtained by crossing the current that we  employed in the  
top quark decays computation~\cite{Brucherseifer:2013iv}. 

In computing radiative corrections to top quark decay we were able to argue that, 
due to  simple kinematics of that process, 
we do not need to consider a true extension of momenta of any particle to $d$-dimensional space.  This is so  because, when  the initial 
top quark is at rest, for all sub-processes it is possible to choose a reference frame where momenta of all particles can be parametrized 
by four-dimensional vectors~\cite{Brucherseifer:2013iv}.  However,  in  the case of single-top production, the kinematics is richer 
and we are forced to extend the parametrization of momenta of final state particles in such a way that explicit $(d-4)$-dimensional 
momenta components appear. 
We note that such an extension 
 is non-trivial if we want to {\it i}) keep the \emph{phase-space locality} of subtraction terms 
and {\it ii}) represent $(d-4)$-dimensional momenta components in such a way that 
the numerical 
integration of amplitudes remains 
 possible.  
To deal with this issue, we closely follow the implementation described in  Ref.~\cite{Boughezal2013}.

\begin{table*}
\centering
\begin{tabular}{|c|c|c|c|c|c|}
\hline
$ p_\perp$ & $\sigma_{\rm LO}$, pb  & $\sigma_{\rm NLO}$, pb & $\delta_{\rm NLO}$& 
 $\sigma_{\rm NNLO}$, pb & $\delta_{\rm NNLO}$ \\
\hline
\hline 
0~{\rm GeV}          & $~53.8^{+3.0}_{-4.3}~$  & $ ~55.1^{+1.6}_{-0.9} $  & $+2.4\%$   & $~54.2^{+0.5}_{-0.2} $
 & $-1.6\%$   \\
20~{\rm GeV}          & $~46.6^{+2.5}_{-3.7}~$  & $~48.9^{+1.2}_{-0.5}  $  & $+4.9\%$  & $~48.3^{+0.3}_{-0.02} $
 & $-1.2\%$ \\
40~{\rm GeV}          & $~33.4^{+1.7}_{-2.5}~$  & $~36.5^{+0.6}_{-0.03} $ &  $+9.3\%$   &$~36.5^{+0.1}_{+0.1}$
  &  $ -0.1\%$ \\
60~{\rm GeV}          & $~22.0^{+1.0}_{-1.5}~$  & $~25.0^{+0.2}_{+0.3} $ & $+13.6\%$  & $~25.4^{-0.1}_{+0.2}$  & 
$ +1.6 \%$ \\
\hline
\end{tabular}
\caption{ QCD corrections to $t$-channel single top quark production 
cross sections at $8~{\rm TeV}$ LHC with a cut on the transverse momentum of the top quark $p_\perp$.
Cross sections are shown at leading, next-to-leading and next-to-next-to-leading order 
in  dependence of the factorization and renormalization scale 
$\mu = m_t$ (central value),  $\mu = 2 m_t$ (upper value) 
and $\mu=m_t/2$ (lower value).  Corrections at NLO and at NNLO (relative to the NLO)  are shown 
in percent for $\mu = m_t$. 
}\label{results_table}
\end{table*}

Even after  all the relevant ingredients for the computation of NNLO QCD radiative corrections 
to weak transition 
on a heavy quark line are put together, the result still contains collinear singularities. These singularities 
are removed by the renormalization of parton distribution functions. Since 
parton distribution functions mix under collinear renormalization, we are  forced to consider single top quark 
production in such  partonic channels as $ gW^* \to t \bar b$ that appears first at next-to-leading 
order and $q W^* \to t \bar b q$ that appears first at NNLO. The calculation of radiative corrections 
to those channels proceeds along the same lines as for $W^* b\to t$; the only difference is that 
virtual corrections, either two- or one-loop,  do not necessarily contribute to those channels. 

Although in our approximation NNLO QCD corrections to the heavy quark line are 
treated as  independent
from corrections to the light quark line, the heavy and the light quark lines 
do influence each other due to  kinematic constraints. 
Indeed,  for computing radiative 
corrections, it is convenient to treat the $t$-channel single-top production as a whole  process 
and parametrize kinematics of the full $ub \to d t$ scattering, rather than the kinematics of the
$W^* b \to t$ transition only.  Therefore, when considering corrections to the heavy quark line 
we would like to parametrize the kinematics of a scattering process where 
 a massive particle and a massless particle are produced 
in the collision of two massless particles, and where no singularities are associated with the massless outgoing 
particle. It is easy to realize that the  phase-space parametrization for this case can be directly borrowed from 
the calculation of the Higgs boson production in association with a jet~\cite{Boughezal2013}.
The corresponding formulas for the  phase-space parametrization relevant for the $ub \to d t$, $u b \to d tg$ and 
$ub \to d tgg$ sub-processes,
as well as a discussion of an appropriate choices of variables relevant for the extraction of singularities
can be found in that reference.  
Using  the language of that paper, we only 
need to consider ``initial-state'' sectors since there are no collinear singularities 
associated with final state particles due to the fact that top quarks are massive. 
All calculations required for initial-state sectors are documented in Ref.~\cite{Boughezal2013} except 
that here we need soft and collinear limits for incoming quarks, rather than gluons, and 
the soft current for a massive particle. This, however,  
is a minor difference that does not affect the principal features of the computational method. 

The above discussion of the NNLO QCD 
 corrections to the heavy quark line  can be applied almost verbatim
to  corrections to the light quark line. The 
two-loop corrections for the $0 \to q \bar q' W^*$ vertex are known since long 
ago~\cite{Kramer:1986sg,Matsuura:1987wt,Matsuura:1988sm}. One-loop corrections 
to  $0\to q \bar q' g W^*$  scattering are also well-known;
we implemented the result presented in~\cite{Garland:2002ak} and again checked the implementation against 
an independent computation based on the 
Passarino-Veltman reduction. Apart from different amplitudes,
the only minor difference with respect to  corrections 
to the heavy quark line  is that in this case there are collinear singularities associated
with both, the incoming and the outgoing quark lines. We deal with this problem splitting the 
real-emission contribution into sectors, see
Ref.~\cite{Boughezal2013}. 
In the language of that paper, we have to consider ``initial-initial'', ``final-final'' 
and mixed ``initial-final'' sectors. 
Finally, we briefly comment on the contribution  shown in Fig.\ref{feynman_blocks}c.
We note that,  although formally NNLO, it is  effectively the 
product of NLO corrections to the heavy and the light quark lines, so that 
it can be dealt with using  techniques familiar from  NLO computations.

We will now  comment on our treatment of $\gamma_5$. For perturbative calculations at higher orders the presence 
of the Dirac matrix $\gamma_5$ is a nuisance since it can not be continued to $d$-dimensions in a straightforward way.
While computationally-efficient ways to deal with 
$\gamma_5$  in computations, that employ dimensional regularization, exist  
(see e.g. Ref.~\cite{larin}), 
they are typically complex and untransparent. Fortunately, there is a simple way to solve the $\gamma_5$ problem in our case. Indeed, 
in the calculation of virtual corrections 
to the $tWb$ weak vertex, $\gamma_5$ is taken to be anti-commuting~\cite{nnlov1,nnlov2,nnlov3,nnlov4}. 
This  enforces the left-handed 
polarization  of the $b$-quark and removes the issue of $\gamma_5$ altogether.  Indeed, if we  imagine that the weak $b \to t$ transition 
is facilitated by the vector current but we select the $b$-quark with left-handed polarization only, we will obtain the same result 
as when the calculation is performed with the anti-commuting $\gamma_5$.
Since the cancellation of infra-red and collinear divergences occurs for each polarization of the 
incoming 
$b$-quark separately,  this approach completely eliminates the need to specify the scheme for dealing with $\gamma_5$ 
and automatically enforces simultaneous conservation of vector and axial currents -- a must-have feature 
if quantum anomalies are neglected. Of course, this requires that we deal with the $\gamma_5$ appearing in real emission diagrams 
in the same way as in the virtual correction and this is, indeed, what we do by using  helicity amplitudes, 
as described in~\cite{Brucherseifer:2013iv}.

\begin{table*}[t]
\centering
\begin{tabular}{|c|c|c|c|c|c|}
\hline
$ p_\perp$ & $\sigma_{\rm LO}$, pb  & $\sigma_{\rm NLO}$, pb & $\delta_{\rm NLO}$& 
 $\sigma_{\rm NNLO}$, pb & $\delta_{\rm NNLO}$ \\
\hline
\hline 
0~{\rm GeV}           & $~29.1^{+1.7}_{-2.4}~$  & $ ~30.1^{+0.9}_{-0.5} $   & $+3.4\%$   
&  $~29.7^{+0.3}_{-0.1}$  & $-1.3\%$    \\
20~{\rm GeV}          & $~24.8^{+1.4}_{-2.0}~$  & $~26.3_{-0.3}^{+0.7}  $  & $+6.0\%$   
& $~26.2^{-0.01}_{-0.1}$    &  $ -0.4\% $
   \\
40~{\rm GeV}          & $~17.1^{+0.9}_{-1.3}~$  & $~19.1^{+0.3}_{+0.1} $   &  $+11.7\%$   
& $~19.3^{-0.2}_{+0.1} $ & $+1.0\%$
   \\
60~{\rm GeV}          & $~10.8^{+0.5}_{-0.7}~$  & $~12.7_{+0.2}^{+0.03} $ & $+17.6\%$  
&  $~12.9^{-0.2}_{+0.2}  $  &  $+1.6\%$ \\ 
\hline
\end{tabular}
\caption{ QCD corrections to the $t$-channel single anti-top  production 
cross sections at $8~{\rm TeV}$  LHC with a cut on the transverse momentum of the anti-top quark $p_\perp$.
Cross sections are shown at leading, next-to-leading and next-to-next-to-leading order 
in  dependence of the factorization and renormalization scale 
$\mu = m_t$ (central value),  $\mu = 2 m_t$ (upper  value) 
and $\mu=m_t/2$ (lower value).  Corrections at NLO and at NNLO (relative to the NLO)  are shown 
in percent for $\mu = m_t$. 
}\label{results_table_atop}
\end{table*}

We have performed several checks to ensure that our calculation of NNLO QCD corrections 
to single top quark production is correct. For example, we have compared all the 
tree-level matrix  elements  that are used  in this computation, e.g. 
$ u b \to d t +ng$, with $0 \le  n \le 2$, $ u b \to d  t + q \bar q $, $u g \to d \bar b t + mg$, $ 0 \le m \le 1$, 
   against  MadGraph~\cite{Alwall:2011uj} and  found complete agreement.  
We have  extracted one-loop amplitudes for $ 0 \to W t \bar b g$ from MCFM~\cite{mcfm} and checked them against our own implementation 
of the Passarino-Veltman reduction, for both the $W^* b \to tg$ and the $W^*g \to t \bar b$ processes.  
We  have cross-checked one-loop 
amplitudes for $W^*u \to dg$ and related channels against MadLoop~\cite{Hirschi:2011pa}.
In the intermediate stages of the computation, we also require reduced tree 
and one-loop amplitudes computed  to  higher
orders in $\ep$, as explained e.g. in Ref.~\cite{Boughezal2013}. 
We checked that their contributions drop out from the final results, in accord with 
the general conclusion of Ref.~\cite{Weinzierl:2011uz}.

One of the most important checks is provided by the  cancellation of infra-red and collinear divergences. 
Indeed, the technique for NNLO QCD computations 
described in Refs.~\cite{Czakon:2010td,Czakon:2011ve,Boughezal:2011jf} leads to a Laurent expansion of different contributions 
to differential cross sections  in the dimensional regularization parameter $\ep$; coefficients 
of this  expansion  are computed by numerical integration. Independence of physical cross sections on the regularization 
parameter is therefore achieved numerically, when different contributions to such cross sections (two-loop virtual 
corrections, one-loop corrections to single real emission contributions, double real emission contributions, 
renormalization, collinear subtractions of parton distribution functions,  
etc.)  are combined.  The numerical cancellation of the ${\cal O}(\ep^{i})$ contributions, 
$ -4 \le i  \le -1$  is an important 
check of the calculation.  We computed  partonic cross sections for the $t$-channel single-top production 
at three different center of mass energies and observed cancellation of $1/\ep^4,1/\ep^3,1/\ep^2$ and $1/\ep$ singularities.
For the $1/\ep$ contributions to the cross section, we find that the cancellation is at the per mill level, 
independent of the center-of-mass collision energy.
 For higher poles, cancellations improve
by, roughly,  an order of magnitude per power of $1/\ep$.  We have also checked that similar degree of cancellations is achieved 
for hadronic cross sections, which are computed by integrating partonic cross sections with parton distribution functions.

\section{Results} 
\label{results} 

We are now in position to present the results of our calculation.  We have chosen to consider the $8~{\rm TeV}$ LHC. 
We  use the MSTW2008 set for parton distribution functions and $\alpha_s$;  
when results for N$^k$LO cross sections are reported, the relevant PDF set and $\alpha_s$ 
value are used. 
 We also set the CKM matrix to the identity matrix, 
the top quark mass to $m_t = 173.2~{\rm GeV}$, the Fermi constant to $G_F = 1.16639 \times 10^{-5}~{\rm GeV}^{-2}$ 
and the mass of the $W$ boson to $80.398~{\rm GeV}$.  The factorization and renormalization scales 
are by default set to the value of the top quark mass $m_t$ and varied by a factor two to probe sensitivity of the 
results to these unphysical scales.\footnote{
We note that by comparing NLO QCD corrections to single-top production in four- and five-flavor 
schemes, it was suggested~\cite{nlo7} that choosing $m_t/2$ as a central value is more 
appropriate. Given  reduced dependence on the renormalization/factorization
scales at NNLO,  this issue is less relevant for our computation.}
We account  for all partonic channels. 
At LO, this means that the light quark transition is initiated either by an up-type
quark or by a down-type anti-quark, while the heavy quark transition can only be initiated by a $b$-quark. At NLO, the gluon channel
opens up, both for the light and the heavy quark transitions. At NNLO, 
in addition to that,  we also have to take into account pure singlet contributions,
for example 
 $W^* b \to b \bar u d$ for the light quark line and $W^* u \to u \bar b t$ for the heavy quark line. 
Although we include all partonic channels in our calculation, it turns out that their contributions to single-top production 
differ significantly.  Indeed, we find that it is important to include 
$bu \to d t$, $gu \to d t \bar b $,  $q  u  \to d  q t \bar b$ and  $gb \to q \bar q' t$ in the computation of NLO and NNLO 
QCD corrections while other channels can, in principle, be neglected. 

The simplest observable to discuss is the total cross section.  
Using the input parameters given in the previous paragraph, we find the leading order  cross section 
for single-top production at $8$ TeV LHC  to be  $\sigma_t^{\rm LO} = 53.8~{\rm pb}$, if we set the renormalization and factorization 
scales to $\mu = m_t$. 
The next-to-leading  order QCD cross section at $\mu = m_t$ is $\sigma_t^{\rm NLO} = 55.1~{\rm pb}$, corresponding   to an increase of the 
leading order cross section by $2.5$ percent.  
It is important
to realize that this small increase is the  result of significant  cancellations between various sources of QCD corrections. 
For example, NLO QCD corrections in the $bq$ partonic channel  increase the leading order 
cross section by $10\%$, which is more in line with the expected size of NLO QCD corrections. 
However,  this positive correction is largely canceled by 
the quark-gluon channel that appears at next-to-leading order  for the first time. The gluon-initiated channels have large and negative 
cross sections. Indeed, 
the  $qg \to t \bar b q'$ and $g b \to q \bar q' t$ partonic processes change the leading 
order cross section by $-14\%$. When the leading order cross section is computed with NLO PDFs, it increases 
by $8\%$. Finally, when all the different contributions are combined, 
a small positive change in the single-top production cross section at NLO is observed. 
The scale dependence 
of leading and next-to-leading order 
cross sections is shown in Table~\ref{results_table}.  For the total single-top production cross section, we observe that 
the residual scale dependence at NLO is at a few percent  level.  For $\mu = m_t$, the NNLO QCD cross section 
is $\sigma_t^{\rm NNLO} = 54.2~{\rm pb}$, corresponding to a decrease of the NLO cross section by $-1.5\%$. 
The magnitude of 
NNLO corrections 
is similar to the NLO corrections, illustrating the accidental smallness of the latter.  As can be seen from
Table~\ref{results_table},  the residual scale dependence  of the NNLO result 
is very small, of the order of one percent.

The simplest observable, beyond the total cross section  that one can study, is the cross section with a cut on the 
transverse momentum of the top quark. The corresponding cross sections and QCD corrections are shown in Table~\ref{results_table}.
It follows from there that   the QCD corrections strongly depend  
on the minimal value of the top quark transverse momentum. As we already mentioned, the total cross section, corresponding 
to $p_\perp = 0$ exhibits very small NLO QCD corrections that are, in fact, comparable to the 
residual scale uncertainty. 
At higher values of the cut on the top transverse momentum, the corrections become larger and reach $+14\%$ at $p_\perp = 60~{\rm GeV}$.
However, the magnitude of the scale uncertainty of the NLO QCD prediction  is nearly $p_\perp$-independent
and remains at a few percent level for both small and large values of $p_\perp$.  This suggests 
that, once the  NNLO QCD corrections are taken into account, the single-top production cross section 
becomes  known with a few percent precision for all values of  $p_\perp$.  This is indeed what happens, 
as one can see from Table~\ref{results_table}. Indeed, regardless of the size of NLO QCD corrections -- that 
are very different for different values of $p_\perp$ -- the NNLO QCD corrections are always in the range of just 
a few percent and the residual scale dependence is also in the one percent range. Therefore, availability of the NNLO QCD corrections 
enables very accurate predictions for single-top production for all values of the top quark transverse momentum.

We can also study the production of single anti-top quarks in proton-proton collisions. The corresponding results for the 
total cross section are shown in Table~\ref{results_table_atop}. The magnitude of QCD corrections  for anti-top 
 are similar to that of the top, although they are somewhat larger. Nevertheless, also for the $\bar t$ case one can see 
an impressive stabilization of the NNLO QCD cross sections and only marginal residual dependence on the factorization and 
 renormalization scales.   

It is common in experimental analyses to measure and quote  the sum of top and anti-top production 
cross sections. The combination of ATLAS and CMS $t$-channel single-$t$ and single-$\bar t$ measurements at the $8$~TeV LHC 
was recently given  in Ref.~\cite{atlasconf2013098}. They find $\sigma_{t \; \& \; \bar t}^{\rm exp} = 85 \pm 12~{\rm pb}$. Our NNLO result 
follows from the sum of relevant entries at first rows in Tables~\ref{results_table} and~\ref{results_table_atop}. We obtain 
$\sigma_{t \; \& \; \bar t}^{\rm NNLO} = 83.9^{+0.8}_{-0.3}~{\rm pb}$, in good   agreement with the measured value. 
For comparison, the NLO result $\sigma_{t \; \& \; \bar t}^{\rm NLO} = 85.2^{+2.5}_{-1.4}~{\rm pb}$ is similar, 
but significantly less precise than the NNLO result.

Another  interesting observable~\cite{atlasconf2012056,Khachatryan:2014iya}
is the  ratio of the single-$t$  and single-$\bar t$ cross sections, since 
this ratio is  sensitive to the relative size  of parton distribution functions for up and down quarks at moderate values of 
the Bjorken variable $x$. 
This ratio also depends on the top (or anti-top)  $p_\perp$ cut. 
For $p_\perp = 0$ we find  $\sigma_{t}/\sigma_{\bar t} = 1.849 \pm 0.005,\;\;1.831 \pm 0.001,\;\;\;
1.825 \pm 0.001$ at leading, next-to-leading and next-to-next-to-leading order; the recent experimental result~\cite{Khachatryan:2014iya} is
$1.95 \pm 0.1 ({\rm stat}) \pm 0.19 ({\rm syst})$. As it often happens with ratios, 
changes caused by the  scale variation  are probably 
not good indicators of uncertainty of theoretical predictions, especially so since 
the QCD corrections are small.  For higher values of $p_\perp$, the ratio of top and anti-top cross sections increases. 
For example, taking  $p_\perp = 60~{\rm GeV}$, we find 
$\sigma_{t}/\sigma_{\bar t} = 2.037 \pm 0.007,\;\;\;1.969 \pm 0.01,\;\;\;
1.969 \pm 0.02$ at leading, next-to-leading and next-to-next-to-leading order, respectively. 
We stress  that in all cases the errors on the ratio do not include the 
PDF uncertainty which should be significant given the 
smallness of scale-variation errors.  Turning this argument around, we note that, 
 since the perturbative uncertainty on the ratio of top and anti-top production 
cross sections is very small, the  precisely measured ratio 
of single-top and single-anti-top cross sections can  be used to provide additional stringent 
constraints on  parton distribution function.

\section{Conclusions}
\label{concl}

In this paper we described the calculation of NNLO QCD corrections to $t$-channel 
single-top production  cross section at the LHC. We found that the NNLO QCD corrections are small, of the order 
of a few percent. The residual scale dependence of the single-top production cross section  is  below one 
 percent,  indicating that uncalculated higher order corrections are very small. Similar conclusions  are 
reached also for the single-top production cross sections with a cut on the top transverse momentum although 
in that case the residual scale uncertainty can be larger. 
It is interesting to note that, with a cut on  the top quark transverse momentum, 
the NLO QCD corrections are very different for different values of $p_\perp$ but  the NNLO 
QCD corrections and residual scale dependences are small, independent of it.  Therefore, 
it appears that, similar to the Drell-Yan processes $pp \to Z, W$ etc. and the $t \bar t $ pair production 
$pp \to t \bar t$, the single-top production cross section at the LHC  is predicted with a percent-level 
accuracy at NNLO QCD.  In principle, this should allow 
very precise  measurements of the top quark electroweak couplings and ensuing indirect determinations 
of the top quark width.   

A natural extension of our calculation is to include decays of the top quark in the narrow width 
approximation. This can be done in a relatively straightforward way 
by calculating  all the scattering amplitudes, 
needed  to describe single-top production,
for a particular top quark spinor, that  effectively accounts for the complete 
decay chain of the top quark, see e.g. Ref.~\cite{msch}. Once this is done, computation of fiducial 
volume cross sections for realistically selected final states in single-top production as well 
as realistic kinematic distributions of top quark decay products 
becomes possible. We hope that such an extension of the present work 
will contribute towards improved measurements  of the top quark electroweak coupling, 
the top quark width and the CKM matrix element $|V_{tb}|$ at the LHC.

{\bf Acknowledgments} 
We are grateful to  E. Mariani, M. Zaro and the authors of the VBF@NNLO code \cite{vb1,vb2}
for providing numerical cross-checks of parts of this computation. 
This research is partially supported by US NSF under grants PHY-1214000. The research of K.M. and
M.B. is partially supported by Karlsruhe Institute of Technology through its distinguished 
researcher fellowship program. The
research of M.B. is partially supported by the DFG through the SFB/TR 9 ``Computational Particle
Physics''. 
Some of the calculations reported in this paper were performed on the Homewood High Performance Cluster 
of the Johns Hopkins University.

\end{document}